\documentclass[final,5p,10pt,times,twocolumn]{elsarticle}
\usepackage{amssymb}
\usepackage{amsmath}
\usepackage{lipsum}
\usepackage{url}
\journal{Chaos, Solitons, \& Fractals}

\begin{document}

\begin{frontmatter}



\title{Ubiquitous order known as chaos}


\author[Ovchinnikov]{Igor V. Ovchinnikov}
\affiliation{organization={R\&D, CSD, ThermoFisher Scientific Inc},%Department and Organization
            addressline={200 Oyster Point}, 
            city={South San Francisco},
            postcode={94080}, 
            state={CA},
            country={USA}}
\begin{abstract}
A close relation has recently emerged between two of the most fundamental concepts in physics and mathematics: chaos and supersymmetry. In striking contrast to the semantics of the word 'chaos,' the true physical essence of this phenomenon now appears to be a spontaneous order associated with the breakdown of the topological supersymmetry (TS) hidden in all stochastic (partial) differential equations, \emph{i.e.}, in all systems from a broad domain ranging from cosmology to nanoscience. Among the low-hanging fruits of this new perspective, which can be called the supersymmetric theory of stochastic dynamics (STS), are theoretical explanations of 1/f noise and self-organized criticality. Central to STS is the physical meaning of TS breaking order parameter (OP). In this paper, we discuss that the OP is a field-theoretic embodiment of the 'butterfly effect' (BE) -- the infinitely long dynamical memory that is definitive of chaos. We stress that the formulation of the corresponding effective theory for the OP would mark the inception of the first consistent physical theory of the BE. Such a theory, potentially a valuable tool in solving chaos-related problems, would parallel the well-established and successful field theoretic descriptions of superconductivity, ferromagentism and other known orders arising from the spontaneous breakdown of various symmetries of nature. 
\end{abstract}
\end{frontmatter}

%\tableofcontents

%\linenumbers


% Please give the surname of the lead author for the running footer

% Please add a significance statement to explain the relevance of your work
%\significancestatement{The extreme sensitivity to initial conditions, known as the butterfly effect, is a definitive feature of the ubiquitous phenomenon called chaos. While it is often perceived simply as a manifestation of the unpredictability of chaotic dynamics, the butterfly effect is essentially an infinitely long memory that, when sufficiently understood, could potentially become a valuable part of the theory of nonlinear dynamics or even be harnessed for applications. Until now, its consistent theory did not exist, but it may soon appear from the supersymmetric theory of stochastic dynamics -- a multidisciplinary construction at the intersection of dynamical systems theory, stochastic dynamics, and high-energy physics theory. In this paper, we discuss the general physical picture of the butterfly effect that follows from this theory.}

\section{Introduction}

It is well known that a single bit error can result in the freezing of the operating system. Given that altering a single bit represents the smallest conceivable perturbation within the realm of digits, the remarkably dramatic impact of such a minute change can be interpreted as a digital version of the renowned 'butterfly effect' (BE). This term is used to describe the extreme sensitivity to perturbations and/or initial conditions in 'chaos' (see, \emph{e.g.}, Refs.\cite{RevModPhys.57.617,Chaos_book_1,Chaos_book_2017,Gilmore,10.1063/5.0025924,Yorke_1975,TransientChaos2015,Intermittency_Review,Baxendale.10.1007/BFb0076851,Arnold.10.1007/BFb0076835} and Refs. therein), a ubiquitous nonlinear dynamical phenomenon with a scientific history spanning over a century \cite{Mot14,Rue14,Poincare_celestial_dynamics,ButterFly}.

Labeling the CPU dynamics as chaotic underscores the idea that, if visualized, the dynamics would indeed appear random to a human observer. From the same observer's perspective, however, the most valuable aspect of this seemingly erratic dynamics is the logic that the CPU is designed to execute. Logic, in turn, is conceptually closer to 'order' -- the polar opposite of 'chaos'. In other words, 'chaos' may not always be the most fitting label for dynamics exhibiting the BE. 

To reach this conclusion, we intentionally stretched the concept of chaos for the sake of argument. In reality, chaos is conventionally a descriptor of dynamics of continuous variables rather than discrete digits. As we discuss next, however, the above conclusion remains valid for dynamics of continuous variables, and even for stochastic dynamics, \emph{i.e.}, dynamics affected by external random noise \cite{Baxendale.10.1007/BFb0076851,Arnold.10.1007/BFb0076835, ARNOLD19831,Oks10,LEJAN1984307,Kunita2019,Hairer_2001,kupiainen2016renormalization,Heirer_2018,Bedrossian_RecentReview}.\footnote{Chaos is often described as a deterministic phenomenon. Nonetheless, deterministic dynamics is a mathematical idealization. Real-world dynamical systems are always influenced by external noise. Therefore, the solid generalization of chaos to stochastic models, as provided by the supersymmetric theory of stochastic dynamics, is an important theoretical advancement in its own right.}

The rationale behind this assertion begins with the observation that all stochastic differential equations (SDE) possess a hidden topological supersymmetry (TS) \cite{Ovc11,OvcEntropy,Torsten,DMM_2}. This symmetry can be viewed as an algebraic manifestation of the smoothness of continuous-time dynamics, a generalization of the N=2 supersymmetry of Langevin SDEs \cite{CG,DH,Dijkgraaf,KS,ZinnJustin,Lyapunov_SUSY}, and a key ingredient of cohomological topological field theories (TFT) \cite{Witten_1982, Witten98, Witten981, Baulieu_1988,Baulieu_1989,labastida1989,Blau,Brooks,TFT_BOOK,Frenkel2007215}. From a physicist's perspective, this TS  exists across a broad domain, ranging from the large-scale structure of the Universe, down through numerous diverse fields such as geology and biology, and all the way to the atomic level. Accordingly, the true physical essence of chaos, or more precisely, its stochastic generalization, is the spontaneous breakdown of this pervasive TS -- a picture that harmoniously aligns with the omnipresence of chaos in the natural world.

In a broader context, \emph{spontaneous} symmetry breaking is a fundamental phenomenon (see, \emph{e.g.}, Ref.\cite{Kible_10.1007/978-94-007-1029-0_1} and Refs therein) defined as a situation where a system's ground state possesses lower symmetry than the system itself.\footnote{This phenomenon can be contrasted with \emph{explicit} symmetry breaking, wherein the symmetry of the system itself, rather than just that of the ground state, is broken by external perturbation, with the Zeeman effect being one of the most recognizable examples from atomic physics.} Its physical essence lies in the spontaneous emergence of 'order', or a qualitatively new property, that explicitly breaks the symmetry of the system. For instance, in contrast to the fluid phase of a substance, its solid phase possesses a lattice structure that breaks the translational symmetry of the equations of motion of its constituent atoms. Another example is a metal: in the ferromagnetic phase, it exhibits spontaneous magnetization breaking the rotational symmetry, while in the superconducting phase, it has a Bose-Einstein condensate of Cooper-pairs breaking the conservation of the number of electrons.

\begin{figure}[t]
\centering
  \includegraphics[width=0.99\linewidth]{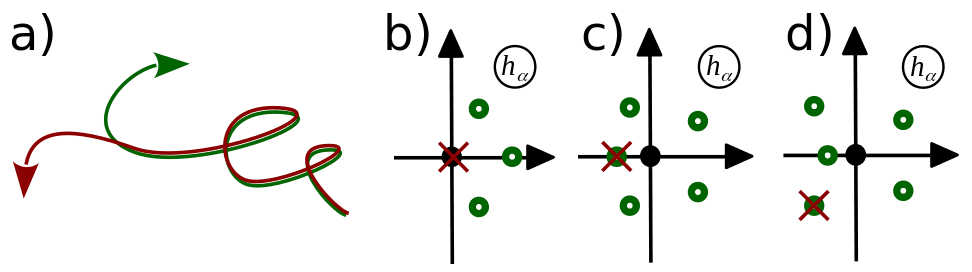}
\caption{{\bf (a)} The qualitative way to see that chaos is the spontaneous breakdown of topological supersymmetry (TS) is to recall that its definitive property -- the butterfly effect -- is the dynamic separation of initially close points, thereby "breaking" the proximity of points or the topology of the phase space. In the algebraic representation of dynamics, the limit of the long evolution corresponds to the ground state, whereas the "breaking of topology" implies that the ground state is not symmetric with respect to TS, which is the spontaneous symmetry breaking by definition. {\bf (b)-(d)} Three possible types of spectra of stochastic evolution operator as discussed in text. In {\bf (b)}, the ground state is the zero-eigenvalue supersymmetric eigenstate (black filled circle), indicating unbroken TS. In {\bf(c)} and {\bf (d)}, the ground states have non-zero eigenvalues, which means that they are non-supersymmetric and TS is spontaneously broken. In {\bf (d)}, the pseudo-time reversal symmetry relating eigenstates with complex conjugate eigenvalues is also broken.}
\label{figure:1}
\end{figure}

In modern theoretical nomenclature, phases with spontaneously broken symmetries are designated as \emph{ordered}, contrasting them with \emph{symmetric} phases featuring unbroken symmetries. Consequently, chaos must be recognized as an ordered phase with the corresponding TS breaking order being the BE (see Fig.\ref{figure:1}a). At this, the BE should not be conventionally perceived as a mere manifestation of the unpredictability of chaotic dynamics; instead, it should be viewed as an infinitely long dynamical memory. Given its infinite range, this memory must inherently exhibit extreme sensitivity to perturbations, akin to the above sensitivity of the dynamics in the CPU to the negation of a single bit. When thoroughly understood, this order may evolve into a valuable component of the theory of nonlinear dynamics or, perhaps, even be harnessed for applications.   

%%%%%%%%%%%%%%%%%%%%%%%%%%%%%%%%%%%%%%%%%%%%
%%%%%%%%%%%%%%%%%%%%%%%%%%%%%%%%%%%%%%%%%%%

This fresh perspective on chaos, coming from what can be called the supersymmetric theory of stochastic dynamics (STS), constitutes a novel and qualitative addition to the well-developed dynamical systems threory, which has already unveiled numerous definitive properties of this fundamental phenomenon. STS seamlessly integrates into the existing chaos framework, as evident particularly from the content of this paper, while simultaneously adding relevance to chaos at its most fundamental level.

For instance, in the original perception of this phenomenon as a form of dynamical mess or disorder -- hence its name 'chaos' --  the assertion that neurodynamics is chaotic typically conjures thoughts of randomness and unpredictability, the qualities impractical for information processing. Within STS, however, the same statement suggests that neurodynamics has an underlying spontaneous order potentially capable of accommodating some form of logic. This understanding adds particular intrigue to the prospect of applying STS to neurodynamics and we will briefly touch upon this subject later in the paper. Our primary focus here, however, is a broader implication of STS.

Namely, despite abundance of literature on chaos, a consistent physical theory of the BE never existed. STS, on the other hand, paves the way toward such a theory, the development of which relies on a deeper understanding of the physical meaning and structure of the TS-breaking order parameter, along with a methodology for construction of the corresponding effective theory. These questions will be addressed in Sec.\ref{section:outlook}, where we argue, among other things, that in some spatially extended systems,\footnote{We use the term a spatially extended system to denote a model defined via a stochastic \emph{partial} differential equations.} BE may admit a holographic description. Let us begin, however, with a brief review\footnote{A more detailed discussion of the theory can be found in Ref.\cite{OvcEntropy}.} of the key elements and results of STS and provide a novel view on wavefunctions and fermion propagators that establish a solid link between STS and dynamical systems theory.

%%%%%%%%%%%%%%%%%%%%%%%%%%%%%%%%%%%%%%%%%%%
\section{Key Elements of STS}
\label{sec:review}
%%%%%%%%%%%%%%%%%%%%%%%%%%%%%%%%%%%%%%%%%%%

The theory began in the late 1970s with the work of Parisi and Sourlas \cite{ParSour,ParSour1}, who introduced the supersymmetric approach to Langevin SDEs \cite{CG,DH,Dijkgraaf,KS,ZinnJustin,Lyapunov_SUSY}. This method was extended to classical mechanics and a few other special classes of SDEs \cite{Gozzi2,Deotto_1,Niemi1,Kurchan}. It was also identified \cite{Baulieu_1988,TFT_BOOK} as a part of the Witten-type or cohomological topological field theories (TFT), the models featured by the presence of TS \cite{Witten_1982, Witten98, Witten981, Baulieu_1988,Baulieu_1989,labastida1989,Blau,Brooks,TFT_BOOK,Frenkel2007215}. More recently, this TS has been established in all SDEs \cite{Ovc11,OvcEntropy,Torsten,DMM_2}, naturally leading to the establishment that chaos, or rather its stochastic generalization, and the spontaneous breakdown of this ever-present TS are one and the same phenomenon. 

%%%%%%%%%%%%%%%%%%%%%%%%%%%%%%%%%%%%%%%%%%%%%%%%%%%%%%%%%%%%%5
%%%%%%%%%%%%%%%%%%%%%%%%%%%%%%%%%%%%%%%%%%%%%%%%%%%%%%%%%%%%%5

\subsection{Wavefunctions and External Observers}
\label{sec:MeaningOfWavefunctions}

%%%%%%%%%%%%%%%%%%%%%%%%%%%%%%%%%%%%%%%%%%%%%%%%%%%%%%%%%%%%%5
%%%%%%%%%%%%%%%%%%%%%%%%%%%%%%%%%%%%%%%%%%%%%%%%%%%%%%%%%%%%%5

The key distinction between STS and the conventional picture of stochastic dynamics (see, e.g., Refs.\cite{Oks10,LEJAN1984307,Kunita2019,Hairer_2001} and Refs. therein) lies in "wavefunctions" -- terminology that we borrow from quantum theory and use loosely.\footnote{Below, additional terminology from quantum (field) theory will be incorporated because not all field-theoretic concepts have counterparts in dynamical systems theory and stochastic dynamics.} In the both cases, the wavefunctions represent external observers.\footnote{This is one of the ways in which stochastic dynamics differs from quantum theory, wherein wavefunctions describe the physical system itself, not the observer.} In the conventional approach, however, the observer is only interested in the original variables of the SDE, which can be expressed in the following general form:\footnote{Summation over repeated indices is assumed throughout the paper.}
\begin{eqnarray}
\dot x= F(x, \xi) = f(x) + e_a(x)\xi^a,\label{SDE}
\end{eqnarray}
where $x\in X$ is the dynamical variable of the system from the topological manifold called the phase space, $X$, $\xi^a, a=1,2...$ is a set of noise variables, and $F$ is a vector field representing the equations of motion with $f$ being its deterministic part and $e$'s being vector fields that couple the noise to the system. 

In the deterministic limit, it suffices to investigate the phase portrait, \emph{i.e.}, the trajectories generated by $f$. In the stochastic case, on the other hand, all trajectories are possible, much like in quantum dynamics. This necessitates probabilistic description that can be achieved particularly by turning to the total probability distribution, $P(x)$. It has the meaning of probabilistic knowledge/ignorance of the external observer about the system and this is the wavefunction of the SDE in the conventional approach.

In STS, on the other hand, the observer is a 'chaotician' with the intentions of investigating the 'group' properties of trajectories, for instance, via Lyapunov exponents. To this end, the observer would define a differential, $dx$, and numerically propagate it along trajectories of SDE according to,
\begin{eqnarray}
d \dot x = (\partial F(x, \xi)/\partial x)  dx.\label{Differentials}
\end{eqnarray}
If multiple Lyapunov exponents are of interest, the observer propagates differential volumes, $dx^{i_1}\wedge ..\wedge dx^{i_k}$, with $\wedge$ being antisymmetric or wedge product. %The information on evolution of symmetric products of differentials is contained in the evolution of the bosonic variables. 
Because of the antisymmetric composition rules, $dx^{i_1}\wedge dx^{i_2}=-dx^{i_2}\wedge dx^{i_1}$, $dx^{i}\wedge dx^{i}=0$, the differentials can be recognized as anticommuting Grassmann numbers or fermions \cite{Witten_1982}. Accordingly, the wavefunction now is a function not only of the original (bosonic) variables but also of these fermions, which we denote as $\chi$: $\chi^{i_1} ... \chi^{i_k} \equiv dx^{i_1}\wedge...\wedge dx^{i_k}$ ($D=Dim X$):
\begin{eqnarray}
\psi(x,\chi) = \sum\nolimits_{k=0}^{D}\psi^{(k)}_{i_i...i_k}(x) \chi^{i_1}...\chi^{i_k}.\label{Eq:DiffForms}
\end{eqnarray}
The right hand side (r.h.s) here is a Taylor expansion in $\chi$ and each term is a differential form -- the coordinate-free object at the core of algebraic topology \cite{Nakahara}. 

In this coordinate-free setting, the total probability distribution is a top differential form, $\psi^{(D)}(x) = P(x)\chi^1...\chi^D$, whereas other wavefunctions of STS are generalizations of this concept to situations with nontrivial fermionic content. They serve as mathematical bookkeeping for the above differentials \cite{Lyapunov_SUSY} suitable for stochastic dynamics when all trajectories are possible.

There may be other perspectives from which to view the wavefunctions in STS. For example, certain class of differential forms can be interpreted as conditional probability distributions \cite{OvcEntropy}. But the exact interpretation of wavefunctions may not be as important\footnote{After all, in quantum theory, wavefunctions do not have a widely accepted meaning.} as the following observation that the above extension of the Hilbert space is physically meaningful. 

Specifically, a consistent theory of stochastic dynamics is expected to encompass a prominent representative of the partition function of the noise, which is a fundamental object that appears at the very formulation of an SDE. And this representative indeed exists, but only if we consider the extended Hilbert space of STS. This representative is the renowned Witten index,
\begin{eqnarray}
W = Tr (-1)^{\hat k}\hat {\mathcal M}_{tt'},\label{eq:WittenIndex}
\end{eqnarray}
where $\hat k$ is the fermion number operator and $\hat {\mathcal M}_{tt'}$ is the stochastic evolution operator (SEO). Up to a topological factor, $W$ equals the partition function of the noise, which is the physical way to understand its topological character \cite{chaos_2}.

%%%%%%%%%%%%%%%%%%%%%%%%%%%%%%%%%%%%%%%%%%%%%%%%%%%%%%%%%%%%%5
%%%%%%%%%%%%%%%%%%%%%%%%%%%%%%%%%%%%%%%%%%%%%%%%%%%%%%%%%%%%%5

\subsection{Stochastic Evolution Operator and Topological Supersymmetry}

%%%%%%%%%%%%%%%%%%%%%%%%%%%%%%%%%%%%%%%%%%%%%%%%%%%%%%%%%%%%%5
%%%%%%%%%%%%%%%%%%%%%%%%%%%%%%%%%%%%%%%%%%%%%%%%%%%%%%%%%%%%%5

Just like the evolution operator in quantum theory, the SEO in Eq.(\ref{eq:WittenIndex}) can be given both pathintegral and operator representations. The pathintegral representation is useful particularly because it allows to easily establish \cite{OvcEntropy} that the presence of TS does not rely on the white noise approximation that we will employ below. It holds true for all SDEs in all interpretations of SDEs, \emph{e.g.}, the Ito interpretation.

At the same time, the pathintegral representation comes with an intrinsic ambiguity known in the domain of stochastic dynamics as the Ito-Stratonovich dilemma. This ambiguity is removed in the operator representation by providing the SEO with its most natural mathematical meaning\footnote{This is actually yet another difference between STS and quantum theory, where the evolution operator does not have a prominent mathematical meaning.} of the generalized transfer operator (see p. 893 of Ref.\cite{Rue02}). 

Namely, for a given noise configuration, an SDE defines a family of noise-configuration-dependent maps of the phase space onto itself. The maps define trajectories as $x(t)=M_{tt'}(x(t'))$. Each map induces an action on wavefunctions or differential forms called pullback,
\begin{eqnarray}
    |\psi(t)\rangle = \hat M^*_{tt'}(\xi) |\psi(t')\rangle, \; \hat M^*_{tt'}(\xi)  = \mathcal{T} e^{- \int_{t'}^td\tau \hat{ \mathcal{L}}_{F(\tau)}},\label{pullback}
\end{eqnarray}
where $\hat{\mathcal L}_{F(\tau)} = \hat{\mathcal L}_{f} + \xi^a(\tau)\hat{\mathcal L}_{e_a}$, is the Lie derivative along the r.h.s of Eq.(\ref{SDE}), $F(\tau)= f + e_a\xi^a(\tau)$, understood as a vector field over the entire $X$ so that the position $x\in X$ must not be specified as an argument. $\mathcal{T}$ denotes chronological ordering. Its essence is evident in,
\begin{eqnarray}
    \hat M^*_{tt'}(\xi)  = \textstyle \hat 1 - \int_{t'}^t d\tau \hat {\mathcal L}_{F(\tau)} + \int_{t'}^t d\tau_1 \int_{\tau_1}^t d\tau_2 \hat{\mathcal L}_{F(\tau_2)}  \hat{\mathcal L}_{F(\tau_1)} + ...,
\end{eqnarray}
which is a formal solution to the operator differential equation $d \hat M^*_{tt'}(\xi)/dt = - \hat {\mathcal L}_{F(t)} \hat M^*_{tt'}(\xi), \left.\hat M^*_{tt'}(\xi)\right|_{t=t'} = \hat 1$, demonstrating that the meaning of the Lie derivative is the infinitesimal pullback. 

The SEO can now be defined as,
\begin{eqnarray}
    \hat {\mathcal M}_{tt'} = %\overline{ \hat M^*_{tt'} } = 
    \overline{\hat M^*_{tt'}(\xi)   },
\end{eqnarray}
where the bar denotes averaging over noise configurations.\footnote{It is worth noting that averaging here is always legitimate because pullback is a linear object, unlike, say, trajectories which cannot be directly averaged over the noise in models with nonlinear phase spaces.}

When the noise variable $\xi$ is gaussian white, $\overline{\xi^a(t)\xi^a(t')}=\delta(t-t')\delta^{ab}$, and dependence of $e$'s on $x$ is nontrivial, Eq.(\ref{SDE}) belongs to the general class of SDEs with multiplicative noise. Its SEO takes a particularly simple form reminiscent of the Matsubara evolution operator in the quantum statistical physics or Wick rotated version of quantum evolution operator,
\begin{eqnarray}
\hat {\mathcal M}_{tt'} = e^{-(t-t')\hat H}, 
\end{eqnarray}
where $\hat H$ is the \emph{infinitesimal} SEO which corresponds to the Stratonovich interpretation of SDEs and/or (bi-graded) Weyl symmetrization rules of quantum theory. 
%%%%%%%%%%%%%%%%%%%%%%%%%%%%%%%%%%%%%%%%%%%%
\begin{figure}[t]
\centering
  \includegraphics[width=0.99\linewidth]{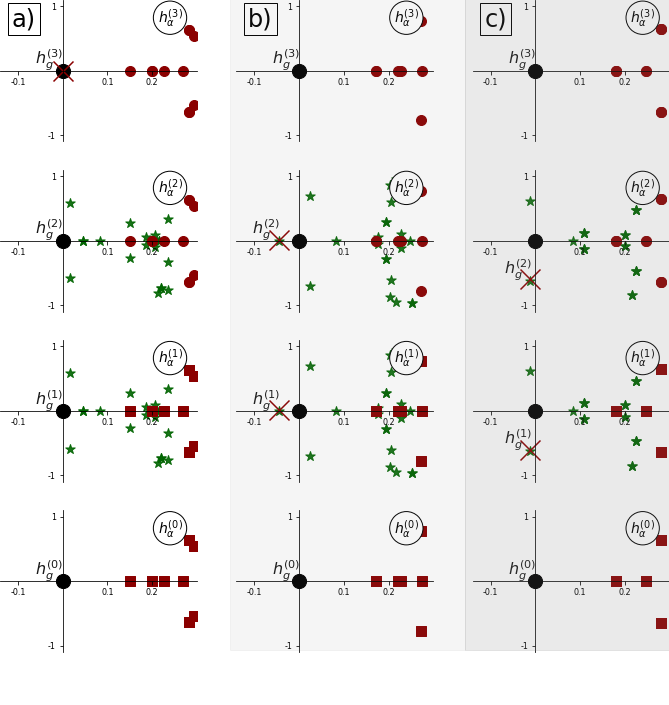}
\caption{Numerical examples of the three types of spectra illustrated in Fig.\ref{figure:1}b-d in stochastic ABC model on a $D=3$ torus. The spectra are presented separately for each degree, from 3 at the top to 0 at the bottom. The parameters of the model are $A=B=1$, $\Theta=0.83$, and $C=0.8, 1.25, 1$ for {\bf a,b}, and {\bf c}, respectively. Each De Rham cohomology class of the phase space provides one supersymmetric eigenstate with zero eigenvalue (black circles at the origin). All other eigenstates are non-supersymmetric pairs that are related by $\hat d$ and have degrees differing by unity. They are presented as red circles, green stars, and red squares for degrees 3 and 2, 2 and 1, and 1 and 0, respectively. Red crosses mark the ground states, which, for cases {\bf b} and {\bf c}, are non-supersymmetric and consequently doubly degenerate. }
\label{figure:2}
\end{figure}
%%%%%%%%%%%%%%%%%%%%%%%%%%%%%%%%%%%%%%%%%%%

The SEO has a very special form definitive for cohomological TFTs: it is $d$-exact, that is, it is a bi-graded commutator of exterior derivative $\hat d=\hat \chi^i\partial/\partial x^i$ with some other operator,
\begin{eqnarray}
\hat H = \hat{\mathcal{L}}_{f} - \hat{\mathcal{L}}_{e_a}\hat{\mathcal{L}}_{e_a}/2 = [\hat{d}, \hat{\bar d}],
\end{eqnarray}
where $\hat{\bar d} = \hat i_f  - \hat i_{e_a} \hat{\mathcal{L}}_{e_a}/2$, where $\hat i_f = i\hat \chi^\dagger_{i} f^i$ denotes the so-called internal multiplication with $\hat \chi^\dagger_i=-i\partial/\partial\chi^i$ being the fermion momentum/annihilation operator, such that $[\hat \chi^i, \hat\chi^\dagger_j] = -i \delta^i_j$, and 
\begin{eqnarray}
    \hat{\mathcal{L}}_{f} = [\hat d, \hat i_f],
\end{eqnarray}
is the Lie derivative given via the Cartan formula that reveals $d$-exactness of this operator. 

Due to the nilpotency property, $\hat d^2=0$, the exterior derivative commutes with any $\hat d$-exact operator including $\hat H$. This commutativity with the SEO suggests that the exterior derivative is the symmetry of the model -- this is the aforementioned TS. 

This symmetry turns the original equations of motion (\ref{SDE}) into the equation of motion for the differentials (\ref{Differentials}). It can be viewed as an algebraic representation of the concept of boundary and as a supersymmetry that adds a fermion to the wavefunction. Furthermore, in the pathintegral representation, 
\begin{eqnarray}
   %\langle x\chi| \hat M_{tt'} |x'\chi'\rangle = \iint e^{ \{Q, \bar Q \} },\label{pathInt}
   \hat{\mathcal M}_{tt'} = \textstyle \int\int%_{ 
   %\stackrel{ \text{paths from } } {x'\chi' \text{at t' to }x\chi \text{ at t}}
   %x\chi(t') = x'\chi', x\chi(t) = x\chi
   %} 
   e^{ - \{Q, \Psi\} } \mathcal{D} x \mathcal{D} \chi \mathcal{D} B \mathcal{D} \bar\chi,\label{pathInt}
\end{eqnarray}
where the functional integration goes over all paths connecting the in- and out- arguments of the SEO and $B,\bar \chi$ are additional fields called Lagrange multipliers, the pathintegral version of TS, $Q=\int d\tau (\chi^i\delta/\delta x^i + B_i \delta/\delta\bar\chi_i)$, is the gauge-fixing Becchi–Rouet–Stora–Tyutin symmetry that restricts the functional integration only to solutions of the SDE. Accordingly, the fermions of STS can be recognized as Paddeev-Popov ghosts \cite{Chen_TFT_doi:10.1142/S0219887813500035,TFT_BOOK} -- fermionic fields that are needed to accomplish the gauge-fixing correctly. 

The entire action in r.h.s. of Eq.(\ref{pathInt}) is $Q$-exact, \emph{i.e.}, the result of acting by $Q$ on some functional called gauge-fermion, $\Psi=\int_{t'}^t d\tau(i\bar\chi_i (\dot x^i - (f^i - e^i_a L^i_{a}/2))$, where $L=\{Q,i\bar\chi_i e^i_a \}$ is the pathintegral version of Lie derivative. Such actions that contain nothing but a gauge-fixing term are definitive for cohomological TFTs \cite{TFT_BOOK}.

The presence of TS is crucial for the entire narrative of STS. But could it be a result of a mistake in derivations ? No, it cannot. The presence of a hidden TS in all SDEs is not just a result of formal manipulations with formulas. It has deep mathematical roots. For example, the $\hat d$-exactness of $\hat H$ is a direct consequence of the $d$-exactness of Lie derivatives. This property is persistent enough to survive averaging over the noise and the SEO simply inherits it from the underlying pullbacks.

Furthermore, TS commutes with the SEO by virtue of its commutativity with the pullback of any differentiable map called diffeomorphism, and SDE-defined maps are diffeomorphisms for all noise configurations \cite{Slavik}. Diffeomorphisms, as such, preserve the proximity of points in the phase space, \emph{i.e.}, the phase space topology. In other words, close initial conditions result in close trajectories. In case of chaotic dynamics, however, this property is 'broken' in the limit of infinitely long evolution (see Fig.\ref{figure:1}a), which is a qualitative way to see that the spontaneous breakdown of TS must be associated with dynamical chaos. On the quantitative level, the spontaneous breakdown of TS is the characteristic of the SEO spectra addressed next.

%%%%%%%%%%%%%%%%%%%%%%%%%%%%%%%%%%%%%%%%%%%%%%%%%%%%%%%%%%%%%%%%%%%55
%%%%%%%%%%%%%%%%%%%%%%%%%%%%%%%%%%%%%%%%%%%%%%%%%%%%%%%%%%%%%%%%%%%55

\subsection{Spectra and Spontaneous Breakdown of Topological Supersymmetry}

%%%%%%%%%%%%%%%%%%%%%%%%%%%%%%%%%%%%%%%%%%%%%%%%%%%%%%%%%%%%%%%%%%%55
%%%%%%%%%%%%%%%%%%%%%%%%%%%%%%%%%%%%%%%%%%%%%%%%%%%%%%%%%%%%%%%%%%%55

SEO has a certain set of general properties \cite{OvcEntropy}.\footnote{For simplicity, we consider only compact finite-dimensional phase spaces and non nondegenerate noises so that the SEO spectra are discrete and the real part of eigenvalues are bounded from below.} For example, SEO is pseudo-Hermitian \cite{Mos02} and its eigenvalues are either real or pairs of complex conjugates.  Due to the presence of TS, all eigenstates are divided into two groups: a few supersymmetric singlets and all the other eigenstates are non-supersymmetric doublets,\footnote{The corresponding bras, or left eigenstates, are related as $\langle \psi_\alpha | = \langle \psi_\alpha' | \hat d$.}
\begin{eqnarray}
    \hat H |\psi_\alpha\rangle = h_{\alpha}|\psi_\alpha\rangle,\hat H |\psi'_\alpha\rangle = h_{\alpha}|\psi'_\alpha\rangle,\; |\psi'_\alpha\rangle = \hat d  |\psi_\alpha\rangle.
\end{eqnarray}
Supersymmetric singlets, $\theta$'s, are in the cohomology of $\hat d$ or in the De Rham cohomology.\footnote{Every De Rham cohomology class must provide one supersymmetric eigenstate. Otherwise, the eigensystem of SEO is not complete which is in contradiction with the theory of pseudo-Hermitian operators.}  That is, 
\begin{eqnarray}
    \hat d |\theta\rangle = 0,\; |\theta\rangle \ne \hat d  |\theta'\rangle, \forall |\theta'\rangle.\label{SUSYStates}
\end{eqnarray}
These states are such that expectation value of any $\hat d$-exact operator vanishes,
\begin{eqnarray}
    \langle \theta | [\hat d, \hat X ] |\theta\rangle = 0, \forall \hat X, 
\end{eqnarray}
including the SEO which is also $\hat d$-exact. This is why the eigenvalue of any supersymmetric state is exactly zero, whereas all eigenstates with non-zero eignevalues are non-supersymmetric. 

Due to commutativity of SEO with the fermion number operator, $\hat k = \chi^i \partial/\partial \chi^i $, each eigenstate has a well defined number of fermions that can also be called the degree of the differential form. In addition, all non-supersymmetric doublets of the top ($D$) and lowest ($0$) degrees have non-negative real parts of their eigenvalues. This last property imposes a constraint: TS cannot be spontaneously broken when the dimensionality of the phase space is below 3. This observation can be regarded as an easy proof of the stochastic generalization of the Poincar\'e–Bendixson theorem \cite{Stochastic_PB_theorem}. Furthermore, this property ensures that when TS is spontaneously broken, the ground state possesses a non-trivial fermionic content. This can be seen as a precursor to the emergence of the Fermi sea of fermions in spatially extended chaotic systems.

The above properties of SEOs limit their spectra to the three types given in Fig.\ref{figure:1}b-d. All three types are realizable as follows particularly from the relation between STS and the theory of the astrophysical phenomenon of kinematic dynamo \cite{Torsten}. It can be numerically demonstrated using the stochastic ABC model \cite{Ovchinnikov_2018}, a toy model for kinematic dynamo. The phase space of the model is a 3-torus and equations of motion are given by Eq.(\ref{SDE}) with
$f=(A\sin z + C\cos y,A\cos z + B\sin x,B\cos x + C\sin y)$, $e^i_a=(2\Theta)^{1/2}\delta_a^i, a=x,y,z$, and $\xi$ being a 3d guassian white noise. The corresponding SEO is $\hat H=\hat{\mathcal L}_{f_{ABC}}+\Theta \hat\bigtriangleup$, where $\hat \bigtriangleup$ is the Laplacian. Spectra of this SEO are exemplified in Fig.\ref{figure:2}.

To see that the the situation with spontaneously broken TS is qualitatively different, let us first point out that there is always the supersymmetric state of the steady-state probability distribution called sometimes 'ergodic zero' or invariant measure, $|0\rangle=P_{ss}(x)\chi^1..\chi^D$. For spectra in Fig.\ref{figure:1}b, this state can be designated as the ground state,  \emph{i.e.}, the state representing the infinitely long unperturbed evolution known as sustained dynamics, which contrasts with transient dynamics.

With its help, one can define the following one-fermion propagator which makes sense for models with linear phase spaces,
\begin{eqnarray}
    G^i_j(t) = \left\{
    \begin{array}{rl}
         i\langle 0 | \hat \chi^i(t)  \hat  \chi^\dagger_j (0) | 0 \rangle,&t>0  \\
         -i\langle 0 | \hat \chi^\dagger_j (0)\hat \chi^i(t) | 0 \rangle,&t<0
    \end{array}
    \right.\label{OneFermionPropagator}
\end{eqnarray}
where $\hat \chi^i(0) \equiv \chi^i$ and $\hat \chi_j^\dagger(t) = e^{t\hat H}\hat \chi_j^\dagger e^{-t\hat H}$ are operators of creation and annihilation of fermions in the Heisenberg representation, and $\langle 0 | =1$ here is the 'bra' of the erdodic zero state. 

The ergodic zero possesses the maximum possible number of fermions. Consequently, any fermion creation operator, $\hat \chi^i$, annihilates it, resulting in $G(t)=0$ for $t<0$. Regarding $t>0$, the magnitude of the propagator can be understood in the traditional sense of dynamical systems theory: if we shift the trajectory by $\delta x^i$ at time $0$, this will induce an average trajectory shift of $G^i_j(t)\delta x^i$ at a later time $t$. %Since the TS remains unbroken, the propagator does not grow in the large $t$ limit, indicating the absence of the butterfly effect.

Turning to the situation with the spontaenously broken TS, let us first point out that the propagation of fermions have a close relation with Lyapunov exponents. It can be revealed by considering the $k$-fermion propagator in the limit, $t\to+\infty$,
\begin{eqnarray}
\langle 0 | \hat O'(t) \hat O(0)| 0 \rangle \sim \text{Re} \langle 0 | \hat O' | g_{D-k} \rangle \langle g_{D-k} | \hat O | 0 \rangle e^{-t h^{(D-k)}_g}. \label{Propagator}
\end{eqnarray}
where $\hat O'$ and $\hat O$ are some operators that create and annihilate $k$ fermions, $g_k$ is the "ground state" among the eigenstates with $k$ fermions and $h^{(k)}_{g}$ is its eigenvalue (see Fig.(\ref{figure:2})). Up to a sign, $h^{(D-k)}_{g}$ equals the sum of $k$ largest stochastic Lyapunov exponents, in the spirit of Pesin formula \cite{Pesin}.

For spectra in Fig.\ref{figure:1}c-d, the leading Lyapunov exponents are positive and fermion propagators diverge exponentially. This implies that 'ergodic zero' is unstable with respect to creation of fermions and the status of the ground state must be passed over to the fastest growing eigenstate with a negative real part of its eigenvalue. Such ground state is non-supersymmetric, which is the definition of the spontaneous breakdown of TS. It must be pointed out that it is pseudo-Hermiticity and the associated complex spectra \cite{Mos02} which allow for the negative real part of eigenvalues. Particularly, Langevin SDEs -- the most popular and the simplest class of SDEs -- have real and non-negative spectra and consequently never break TS spontaneously.

In contrast to the 'ergodic zero', a non-supersymmetric ground state contains fermionic 'holes', causing the one-fermion propagator (\ref{OneFermionPropagator}) to no longer vanish for $t<0$. Using the standard lore of quantum theory, it could be stated that the 'holes' propagate backward in time. This scenario can be reinterpreted from the perspective of dynamical systems as follows: the trajectory shift $\delta x$ at time zero can be considered a consequence of the trajectory shift $G^i_j(t)\delta x^i$ at an earlier time $t<0$.

As it was already mentioned, the ground states represent "sustained dynamics" of the system. In the deterministic limit, it is essentially dynamics on attractors. Accordingly, the wavefunctions of the ground state \cite{OvcEntropy} are the so called Poincare duals\footnote{The Poincare dual, $\psi_{m}$, of a submanifold, $m\in X$, is a differential form which has the property that $\int_m \phi = \int_{X} \psi_m \wedge \phi$, $\forall \phi$. It is a delta-function-like distribution on $m$ and it has differentials in transverse directions.} of attractors (and the global unstable manifolds more generally). For integrable or non-chaotic deterministic flows,\footnote{Non-integrability in the sense of dynamical systems is a mathematical term for deterministic chaos.} attractors are well-defined topological manifolds that can be represented by supersymmetric ground states. 

For chaotic deterministic flows, on the other hand, the attractors are not topological manifolds. They are fractal objects called strange attractors with the backbone structure described by branched unstable manifolds of the topological theory of chaos \cite{Gilmore,10.1063/5.0025924}. Such attractors cannot be represented by supersymmetric ground states, which is yet another qualitative way to see that the spontaneous breakdown of TS is indeed the mathematical essence of chaos.

To quantify this observation in a fully stochastic scenario, we note that the dynamical partition function, $Z_{tt'} = \text{Tr} \hat{\mathcal  M}_{tt'}$, grows exponentially in the limit of long evolution,
\begin{eqnarray}
&\left. Z_{tt'} \right|_{t-t'\to\infty} \propto e^{|Re h_g| (t-t')},    \label{ExpGrowth}
\end{eqnarray}
where $h_g$ is the eigenvalue of the ground state, associated with dynamical pressure \cite{Rue02} and dynamical or topological entropy \cite{Review_Top_Entropy,TopEntropy}. Eq.(\ref{ExpGrowth}) signifies exponential growth of the number of closed trajectories -- the key feature of chaotic behavior. This convincingly demonstrates that the stochastic generalization of chaos is, indeed, the spontaneous breakdown of TS.

As a final remark in this section, we would like to point out that pseudo-Hermitian SEOs possess a pseudo-time-reversal symmetry that pairs up eigenstates with complex conjugate eigenvalues \cite{Mos02}. This symmetry must be broken for the spectrum in Fig.\ref{figure:1}d because only one of such conjugate eigenstates can be designated as the ground state. This situation may bear physical significance, which, for the time being, appears to be one of the unturned stones of the theory.

%%%%%%%%%%%%%%%%%%%%%%%%%%%%%%%%%%%%%%%%%%%%
\begin{figure}[t]
\centering
  \includegraphics[width=0.99\linewidth]{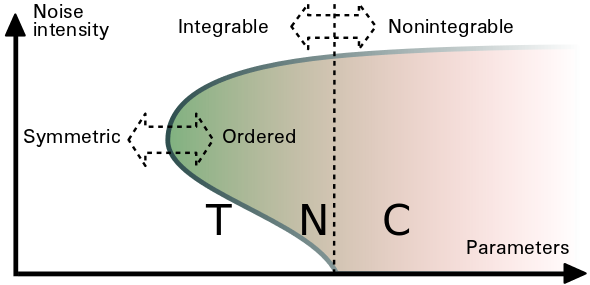}
\caption{Between the symmetric phase ($T$) with unbroken TS and the conventional chaos ($C$), where TS is broken by the nonintegrability of the constant part of the equations of motion ($f$ in Eq.(\ref{SDE})), there exists the phase of noise-induced or intermittent chaos ($N$). In this phase, TS is broken by noise-induced instantons -- the 'quanta' of strongly nonlinear transient dynamics such as earthquakes, neuroavalanches, etc. In the deterministic limit, the $N$-phase collapses into the critical $T$-to-$C$ transition.}
\label{figure:3}
\end{figure}
%%%%%%%%%%%%%%%%%%%%%%%%%%%%%%%%%%%%%%%%%%%%

%%%%%%%%%%%%%%%%%%%%%%%%%%%%%%%%%%%%%%%%%%%%
\begin{figure*}[t]
\centering
  \includegraphics[width=0.95\linewidth]{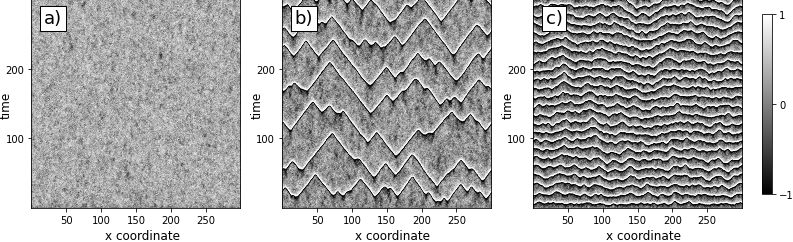}
\caption{Patterns of dynamics (of $\sin\varphi(r,t)$) in the overdamped sine-Gordon equation with non-potential driving on a spatial circle of length 300, which can be viewed as a coarse-grained chain of type-I neurons. In the deterministic limit, the T-C transition (see Fig.\ref{figure:3}) occurs at $\alpha=1$, when vacuum at $\varphi = \sin^{-1}\alpha$ loses stability. Accordingly, Figs. {\bf (a)-(c)}, with parameters $\alpha= 0.92, 0.97$, and $1.1$ (and $\Theta=0.03$), correspond respectively to the $T-$, $N-$, and $C-$ phases as evident from Fig.\ref{figure:5}a. The dynamics in the N-phase {\bf (b)} is dominated by noise-induced instantonic processes of creation and annihilation of pairs of solitons which are the left/right moving kinks/antikinks. In the context of neurodynamics, the kink-antikink pairs can be seen as one-dimensional predecessors of neuroavalanches.}
\label{figure:4}
\end{figure*}
%%%%%%%%%%%%%%%%%%%%%%%%%%%%%%%%%%%%%%%%%%%

%%%%%%%%%%%%%%%%%%%%%%%%%%%%%%%%%%%%%%%%%%%%%%%%%%%%%%%%%%%%%%%%%%%%%%%%%%%%%%%%%
%%%%%%%%%%%%%%%%%%%%%%%%%%%%%%%%%%%%%%%%%%%%%%%%%%%%%%%%%%%%%%%%%%%%%%%%%%%%%%%%%
\subsection{Topological Sector and Instantons}
%%%%%%%%%%%%%%%%%%%%%%%%%%%%%%%%%%%%%%%%%%%%%%%%%%%%%%%%%%%%%%%%%%%%%%%%%%%%%%%%%
%%%%%%%%%%%%%%%%%%%%%%%%%%%%%%%%%%%%%%%%%%%%%%%%%%%%%%%%%%%%%%%%%%%%%%%%%%%%%%%%%

As a TFT, STS encompasses objects that are topological invariants. Among them is the Witten index (\ref{eq:WittenIndex}) and the following matrix elements related to the concept of De Rham cohomology ring \cite{Nakahara}: $\langle \theta_{out}| \hat h_{k}(t_k) ... \hat h_{1}(t_1)|\theta_{in}\rangle$, where $\theta$'s are the global supersymmetric states from Eq.(\ref{SUSYStates}) and $\hat h_{i}(t)$ is the Heisenberg operator constructed from a differential form, $h_i$, nontrivial in De Rham cohomology.

One class of topological invariants may be particularly interesting in the context of dynamics in the noise-induced chaos discussed in the next section. This class (see, e.g., Ref.\cite{Frenkel2007215} and Refs. therein) is the reason why cohomological TFTs are sometimes identified as intersection theory on the so-called instantons (see, e.g., Sec. 10.5 of Ref.\cite{MirrorSymmetry}). 

From the physical point of view, instantons are 'quanta', so to speak, of strongly nonlinear \emph{transient} dynamics, whose examples in nature are earthquakes, lighting bolts, solar flares, neuroavalanches, and many others, including the computational concept of dynamical annealers \cite{DMM_2}. In spatially extended systems, instantons are the processes of annihilation of solitons, such as the kink-antikink annihilation in the $N$-phase of the overdamped sine-Gordon model in the next Sec. \ref{sec:outcomes} (see Fig.\ref{figure:4}b). From theoretical point of view, instantons are families of deterministic trajectories that connect two critical points, or critical manifolds more generally, of the deterministic flow. These families can be defined as intersection of the so-called local unstable and stable manifolds of the two critical points as in Eq.(\ref{instaton}) below.

The following "instantonic" matrix elements are independent of $t$'s ($t_k>...>t_1$) in the deterministic limit:
\begin{eqnarray}
    \langle b | \hat O_k(t_k) ... \hat O_1(t_1) | a \rangle \approx \int_{S_b \cap U_a } O_k\wedge  ...\wedge  O_1.\label{instaton}
\end{eqnarray}
Here, the in- and out- states are the \emph{local} supersymmetric states associated with two critical points of $f$, $a$ and $b$. The states are the Poincare duals of the local unstable ($U_a$) and stable ($S_b$) manifolds. $O$'s are some differential forms and $\hat O$'s are the corresponding operators in the Heisenberg representation.

The discussion in the next section also makes use of the notion of antiinstantons, which are the time-reversed counterparts of instantons.\footnote{For example, in the $N$-phase of the overdamped sine-Gordon model in Sec.\ref{sec:outcomes}, antiinstantons are the noise-induced processes of creation of kink-antikink pairs (see Fig.\ref{figure:4}b).} Their matrix elements contain exponentially small Gibbs factors that vanish in the deterministic limit. This reflects that antiinstantons is essentially dynamics against $f$ in Eq.(\ref{SDE}). Hence, unlike instantons, they can only be induced by noise, and they disappear in the deterministic limit.

%%%%%%%%%%%%%%%%%%%%%%%%%%%%%%%%%%%%%%%%%%%%%%%%%%%%%%%%%%%%%%%%%%%%%%%%%%%%%%%%%
\section{Key results from STS}
\label{sec:outcomes}
%%%%%%%%%%%%%%%%%%%%%%%%%%%%%%%%%%%%%%%%%%%%%%%%%%%%%%%%%%%%%%%%%%%%%%%%%%%%%%%%%
Turning to the physical implications from STS, let us recall first that chaos has a celebrated experimental signature. It is known as 1/f noise, its essence is the long-range dynamical correlations, and it proved to be very resilient to theoretical explanation in the past. 

Within STS, on the other hand, it can be understood as a consequence of the Goldstone theorem. This theorem states that under the condition of the spontaneous breakdown of a global continuous symmetry, there must exist a gapless excitation called the Goldstone particle, or goldstino in the case of fermionic symmetry as in STS. The gaplessness of goldstinos implies that they are long-ranged excitations. By supersymmetry,\footnote{There are infinitely many one-goldstino eigenstates of the Hamiltonian and they are not supersymmetric. Hence, each such state has a bosonic counterpart.} there must also exist a branch of gapless bosonic excitations, which should dominate the long-range response and result in 1/f noise. Alternatively, one can think that integrating goldstinos out would imprint their long-range character onto the bosonic correlations observed in experiments.

Another interesting outcome from STS is the basic phase diagram (see Fig.\ref{figure:3}) that follows from the classification of the analytical levels of TS breaking. In general, there are three levels of spontaneous breakdown of any symmetry: the Gaussian level, by anomaly, which is by perturbatve corrections, and due to noise-induced instantons.\footnote{A more accurate term for noise-induced instantons is a combinations of instantons and noise-induced antiinstantons initiating those instantons.} Supersymmetries, however, are hard to break by anomalies \cite{DynSusyBrWitten}. Accordingly, in most cases there are only two major types of chaos. The first is the conventional chaos, or the C-phase, where TS is broken on the Gaussian level by the non-integrability of $f$ in Eq.(\ref{SDE}). The second type is the noise-induced chaos, or the N-phase, where $f$ is still integrable, but close to being non-integrable, and the noise-induced instantons help break the TS.

The physical picture of dynamics in the $N$-phase is an endless sequence of noise-induced instantons. While these processes can also exist in the $T$-phase as well, it is within the $N$-phase that the noise-induced instantons reveal long-range characteristics, the most prominent of which being the power-law statistics as that for earthquakes on the Richter scale. Naturally, as the system approaches the deterministic limit, the noise-induced instantons vanish, and the $N$-phase collapses onto the critical border of deterministic chaos. 

The $N$-phase is known in the literature \cite{Crutchfield,CRUTCHFIELD200331} as dynamics between 'order and chaos' with a prominent role of 'patterns' - a more common term for solitons/instantons. It is partly covered theoretically by the concept of intermittent chaos (see, \emph{e.g.}, Ref.\cite{Intermittency_Review} and Refs. therein). It is also common to look at the $N$-phase through the prism of self-organized criticality (SOC), \cite{Bak87,ChialvoLoh, SOC_controvercy} a widespread belief that some stochastic models possess a mysterious ability to fine-tune themselves into the critical phase transition into chaos. 

It was proposed as a phenomenological framework to explain 1/f noise in stochastic dynamics dominated by instantons. By its most common formulation, SOC must be perceived as a form of transient dynamics leading to a critical ground state. While it is true that transient dynamics leads to the ground state, the essential question is why the criticality of the ground state within SOC is not accidental, as in normal criticality, but systematic in order to reflect the pervasiveness of 1/f noise. The point is that the properties of the ground state are uniquely determined by the system's position on the phase diagram, which is the space of external parameters that the system cannot change on its own. \footnote{An attempt to reinterpret this autonomous fine-tuning into criticality as a result of the renormalization group (RG) flow, rather than real-time evolution, would inevitably lead to the conclusion that SOC must be associated with attractive critical points of RG to explain the omnipresence of 1/f noise. However, unlike unstable critical points of RG, attractive critical points do not represent critical states with scale-invariant behavior.}
Hence, there is a certain degree of controversy surrounding the idea of autonomous fine-tuning in SOC \cite{SOC_controvercy}.

STS clarifies the physical picture of the $N$-phase identifying it as a full-fledged phase albeit with a diminishing width in the deterministic limit. Furthermore, it reveals that 1/f noise in this phase arises not from a mysterious criticality but rather as a consequence of the Goldstone theorem. This clarification may prove particularly fruitful in neurodynamics, which is dominated by neuroavalanches \cite{beggs2004} and operate in the N-phase \cite{Levina1}. 

To exemplify the three basic phases, we numerically investigated the weak noise limit of an overdamped stochastic sine-Gordon equation \cite{sine_Gordon_Kink_Bath,Classics,Josephson_general,PhysRevB.54.1234}, or Frenkel-Kontorova equation \cite{Review_Frenkel_Kontorova}, with a non-potential driving:
\begin{eqnarray}
\partial_t\varphi = \alpha + \partial_r^2\varphi - \sin\varphi + \sqrt{2\Theta}\xi,    \label{SGE}
\end{eqnarray}
where $\varphi$ is a field on a 2D space $x=(r,t)$, $\alpha$ is the non-potential driving, and $\overline{\xi(x)\xi(x')} = \delta^2(x-x')$ is Guassian white noise. 

This model may be applicable to a number of systems including the voltage dynamics in 1D chains of Josephson junctions \cite{PhysRevB.54.1234}. It was also recognized \cite{ovchinnikov2020} as a coarse-grained version of a 1D chain of type-I neurons (see Sec.I.4.4 of Ref.\cite{NeuronalDynamicsBook}), \emph{i.e.}, a simpler type of neurons that operate at the vicinity of the saddle-node (SN) bifurcation \cite{Izhikevich}, where a point attractor (the resting state) bifurcates into a limit cycle (neuron firing). 

\begin{figure}[t]
\centering
  \includegraphics[width=0.99\linewidth]{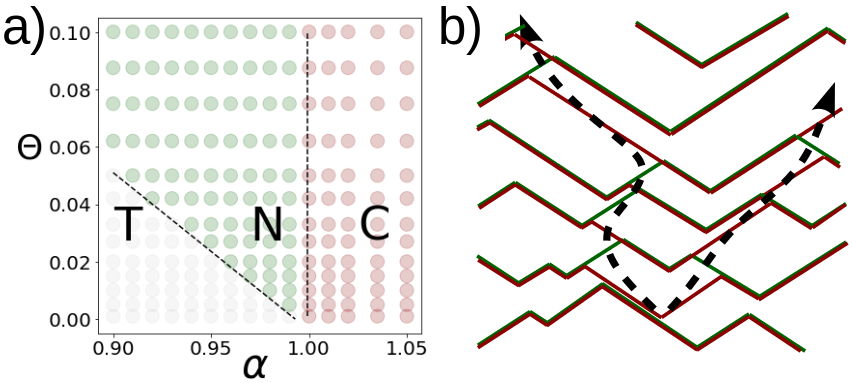}
\caption{{\bf (a)} The part of the phase diagram of the overdamped sine-Gordon equation with non-potential driving Eq.(\ref{SGE}), constructed by maximal stochastic Lyapunov exponents. Grey and colored dots represent negative and positive exponents, respectively. {\bf (b)} A qualitative illustration of the essence of the butterfly effect in the $N$-phase of the model. Near and slightly below $\alpha=1$, even a weak perturbation can introduce a kink-antikink pair, which, due to interaction with other kink and antikinks, forms two "quasi-particles" (dashed curves) of the difference between the perturbed (red) and unperturbed (green) dynamical patterns. These quasi-particles can live for an infinitely long time, thereby manifesting the presence of the butterfly effect.}
\label{figure:5}
\end{figure}

Fig.\ref{figure:4} shows dynamical patterns typical to the three basic phases. The distinctly instatonic character of dynamics in the $N$-phase is evident in Fig.(\ref{figure:4}b). The correspondence between the sets of parameters in Fig.(\ref{figure:4}) and the three basic phases is confirmed by the phase diagram in Figure \ref{figure:5}. The phase diagram is numerically constructed from stochastic Lyapunov exponents \cite{ARNOLD19831,Baxendale.10.1007/BFb0076851,Arnold.10.1007/BFb0076835}. As they should, the exponents are positive in the N-phase, thereby manifesting the presence of the butterfly effect.

In fact, there is a qualitative way to understand the essence of the butterfly effect in the N-phase of this model (see Fig.(\ref{figure:5}b). At and slightly below $\alpha=1$, the system is nearly unstable with respect to creation of kink-antikink pairs. Even a weak perturbation can introduce such a pair, which, upon interaction with other kink and antikinks, forms two ”quasi-particles” of the difference between the perturbed and unperturbed dynamical patterns. These quasi-particles can persist arbitrary long.

We also like to point out that the numerical results from this model were found to qualitatively agree with the corresponding results from neuromorphic hardware \cite{ovchinnikov2020} and clinical data \cite{li2018}.%, and it was speculated that in the context of neurodynamics the three basic phases can be roughly interpreted as coma- (T), consciousness- (N), and seizure-like (C) phases.

%The chaoticity of the $N$-phase was established there by positivity of numerically obtained stochastic Lyapunov exponents, but it can also be seen qualitatively as in Fig.\ref{figure:5}. 
%%%%%%%%%%%%%%%%%%%%%%%%%%%%%%%%%%%%%%%%%%%%
%\begin{figure}[t]
%\centering
%  \includegraphics[width=0.6\linewidth]{Fig_5.png}
%\caption{Comparison of two dynamical patterns of the stochastic overdamped sine-Gordon equation in the N-phase. One pattern (red) is obtained from the other (green) by a small delay (black arrow) of the antiinstantonic creation of the kink-antikink pair at the bottom of the figure. The delay can be though to result from a small perturbation on top of the noise configuration common for the both patterns. The distinction between the patterns caused by the perturbation consists of two "quasiparticles" (depicted by dashed curves) marking the ends of the region where the patterns differ by $2\pi$. Depending on the pattern, these quasiparticles may persist indefinitely, thereby indicating the presence of the butterfly effect.}
%\label{figure:5}
%\end{figure}
%%%%%%%%%%%%%%%%%%%%%%%%%%%%%%%%%%%%%%%%%%%

%%%%%%%%%%%%%%%%%%%%%%%%%%%%%%%%%%%%%%%%%%%%%%%%%%%%%%%%%%%%%%%%%%%%%%%%%%%%%%%%%
\section{Outlook}
\label{section:outlook}
%%%%%%%%%%%%%%%%%%%%%%%%%%%%%%%%%%%%%%%%%%%%%%%%%%%%%%%%%%%%%%%%%%%%%%%%%%%%%%%%%

There are a number of ways in which STS can be advanced further and one of the most promising directions is the development of the methodology for construction of the low-energy effective theory (ET), a simplified description of only important aspects of dynamics represented by the classical object called the TS breaking order parameter (OP).\footnote{In this section we speak only of spatially extended stochastic systems. Similarly to quantum field theory on a lattice, these models can be given the formulation discussed in Sec.\ref{sec:review} by introducing a lattice representation in spatial dimensions.} It offers the prospect of formulating a consistent field-theoretic description of the BE, akin to the well established and successful theories of ferromagnetism, superconductivity \emph{etc.} Given that the ET approach is case-specific and even not unique for a problem at hand, our intention here is to speak only about the most general attributes of ET of the TS breaking OP.

The spontaneous breakdown of a supersymmetry is associated with the condensation of fermions into the ground state. In STS, these fermions, or goldstinos, epitomize unstable directions in the phase space, where the system exhibits the BE \cite{Kang}. Consequently, goldstinos emerge as fundamental constituents for the TS breaking OP, serving as a field-theoretic embodiment of the BE. Given that fermions are typically found in the realm of quantum physics, the situation where the OP of a classical phenomenon is based on fermions may appear somewhat unorthodox.

To address this situation, lets recall that traditional numerical detection of the BE involves taking two copies of the system with close initial conditions and evolving them in parallel over time. In real (not numerical) experiments, however, in most cases (weather, brain, Universe ...) only one copy of the system is available and the noise configuration is not reproducible, which makes the direct measurement of the BE unfeasible. In fact, extracting information about the BE from experimental data requires intricate numerical tools that are a subject of active research \cite{toker2020}. In other words, the BE is a classical phenomenon that cannot be directly measured in experiments. Outside the quantum domain, concepts with such property are rare, if known at all, and it is probably reasonable to assert that the BE is just as bizarre as some purely quantum concepts. In this context, the fermionic character of the corresponding OP no longer appears overly suspicious.

In the Ginzburg-Landau approach, the OP is an expectation value of some local operator. The OP, along with the low-energy ET governing its dynamics, can be introduced in a standard way using the concept of the generating functional,
\begin{eqnarray}
    Z(\zeta) = \langle g |  \mathcal{T} e^{\int_{-T/2}^{T/2}dt \left(\int \zeta^a \hat O_{a} dr - \hat H\right)}| g \rangle,
\end{eqnarray}
where $T$ is the time of evolution, and $\zeta$'s are external "probing" fields that are coupled to a set of local operators $\hat O_{a}$, which is assumed complete is a sense of containing enough data for accurate description of the phenomenon under consideration. 

The action of the ET, $S_\text{eff}=\int_{B} L_\text{eff}$, with $L_\text{eff}$ being the corresponding Langrangian density and $\int_B$ denoting integration over the space-time called the basespace, $B$, can be formally introduced as a functional Legandre transform,
\begin{eqnarray}
\log Z(\zeta)/Z(0) = \min_{\tilde O}\int_{B} \left(\zeta^a \tilde O_{a} - L_\text{eff}(\tilde O) \right),\label{Seff}
\end{eqnarray}
where the limit $T\to\infty$ is assumed and $\tilde O_{a}$ are the OPs that must be viewed as the expectation values of operators $\hat O_{a}$'s.

In the simplest type of TS breaking, which is likely to be realizable in the bulk of $C$-phase, such as the fully developed turbulence in hydrodynamics, the physical picture is the weakly interacting goldstinos forming a Fermi sea. In this situation, OPs can be defined using one-fermion operators. In contrast, dynamics in the $N$-phase is dominated by (anti)instantonic processes each (creating)destroying a tandem of goldstinos. A more relevant picture here would be the condensation of these tandems and the corresponding OP should be based on many-fermion operators. The physical picture would be somewhat similar to superconductivity, where electrons can go in and out of the Bose-Einstein condensate in (Cooper) pairs. 

In formulation (\ref{Seff}), the ET is essentially a classical field theory providing a description of the most interesting part of dynamics of the original quantum-field-theory-like model. Alternatively, one can view the low-energy ET as a result of integrating out the fast degrees of freedom in the spirit of the Willsonain renormalization group approach (see, \emph{e.g.}, Ref.\cite{Kopietz2010} and Refs therein). Similarly to quantum field theories, this procedure may encounter renormalizability issues due to ultraviolet divergences \cite{kupiainen2016renormalization,Heirer_2018}. However, supersymmetries tend to weaken such divergences \cite{SusyReview_2014}. Hence, STS may offer yet another, more practical advantage over the traditional approach to stochastic dynamics, thanks to the presence of the TS.

It must be pointed out that integrating out fast degrees of freedom does not destroy $Q$-exactness of the action. Therefore, the low-energy action must also be $Q$-exact, $S_\text{eff}=\{Q, \Psi_\text{eff}\}$. The same idea can be alternatively expressed by saying that the phenomenon of spontaneous breakdown of a symmetry does not mean that the original symmetry is gone. It only means that the slow part of dynamics becomes qualitatively different particularly due to condensation of OPs. The overall symmetry, which must also involve the OPs, must still be present. In case of STS, this means that OPs must appear in pairs related by the TS, $\tilde O$ and $\tilde O'= \{Q,\tilde O\}$, where $Q$ denotes the pathintegral version of TS as discussed after Eq.(\ref{pathInt}).

%One of the challenges in formulating ET lies in interpreting OPs. This issue is related to the meaning of the fermions in the wavefunctions in section \ref{sec:MeaningOfWavefunctions}. While the current understanding may suffice, it is also possible to remain a topic for further contemplation.

Another interesting point is that the choice of OP is actually not unique. As long as the expectation value of an operator is non-zero only in the ordered phase, it can be designated as the OP, and in case of STS, this holds true for any $d$-exact operator including the SEO. Particularly, the operator must not necessarily be local as in the Ginzburg-Landau approach. Being TFTs, some STSs should have interesting classes of nonlocal operators such as those discussed in Sec. 4.5 of Ref.\cite{TFT_BOOK}. If the ET for the OP based on such operators was established, the resulting picture would reveal strongly nonlocal character of BE making it relative to topological quantum orders (see, \emph{e.g.}, Ref.\cite{topological_order_review} and Refs therein).

%As a final remark, we would like to speculate that the effective theory for the model in Fig.\ref{figure:4} may well be some supersymmetric extension of conformal field theory, where conformality is essential to uphold the scale invariance in the long-wavelength limit stemming from gaplessness of goldstinos\footnote{In comparison to the C-phase, the N-phase must incorporate additional relevant fields (or highest weight states) representing condensed tandems of goldstinos.}. Furthermore, if such an effective theory were subject to AdS-CFT duality \cite{AdSCFT_book}, it would imply the holographic character of the BE. This speculation may be particularly promising in the context of neurodynamics as there are good reasons to believe \cite{Ovchinnikov_2018} that the BE is a 'vessel', so to speak, for the neurodynamical short-term memory and consciousness. Moreover, holography, understood broadly, has long been suspected to be related to the principles for brain function \cite{holography} by analogy with optical holography which is the best known physical realization of associative memory (see, e.g., Ref.\cite{Kohonen1989}).

As a final remark, we would like to speculate that, owing to the gaplessness of goldstinos, some ETs may exhibit scale or even conformal invariance.\footnote{One supporting evidence here could be the scaling properties of the distribution of Lyapunov exponents in a spatially extended system \cite{Yamada_Ohkitani_doi:10.1143/JPSJ.56.4210}.} In this case, the ET would be some supersymmetric and non-unitary extension of conformal field theory (CFT). In fact, a CFT description of some self-organized critical systems has already been proposed \cite{CFT_SOC}. It is worth noting again, however, that the scale-invariance and the corresponding applicability of CFT is not rooted in a mysterious criticality as often mistakenly believed, but rather in the Goldstone theorem.

Furthermore, certain CFT ETs may possess holographic duals, \emph{e.g.}, via AdS-CFT duality \cite{AdSCFT_book}. This framework provides a CFT with an effective classical field theory description in a plus+one dimensional curved basespace called the anti-de Sitter (AdS) space.\footnote{It is often emphasised that the new basespace must only asymptotically be the AdS in the large scale limit.} The additional coordinate, beyond the original space and time, represents the scale of observation, $\mu$. Employing the previously introduced notations, the dual ET can be formally defined as
\begin{eqnarray}
\log Z(\zeta)/Z(0) = \min_{\tilde O, \left.\tilde O_{a}\right|_{\partial AdS}=\zeta^a} \int_{AdS} L_\text{hol}(\tilde O).\label{Shol}
\end{eqnarray}
Here, $\tilde O_{a}$ are classical fields or OPs similar to the ones in Eq.(\ref{Seff}) and the probing fields serve as the conditions for OPs on the boundary of the AdS basespace.\footnote{To be more accurate, the probing fields are the $\mu\to0$ boundary conditions only up to the factors that are algebraic functions of the scale coordinate.}

A holographic description of BE would be particularly interesting in the context of neurodynamics because of compelling reasons to believe that neurodynamical BE may play an important role in the short-term memory and, hence, consciousness \cite{ovchinnikov2020}. Moreover, holography, broadly understood, has long been suspected to be related to the principles of recall \cite{holography}. This follows from analogy with optical holography which is the best known physical realization of associative memory (see, \emph{e.g.}, Ref.\cite{Kohonen1989}), \emph{i.e.}, the key to error-tolerant retrieval \cite{Sommer2012}.
%%%%%%%%%%%%%%%%%%%%%%%%%%%%%%%%%%%%%%%%%%%%%%%%%%%%%%%%%%%%%%%%%%%%%%%%%%%%%%%%%
\section*{Conclusion}
%%%%%%%%%%%%%%%%%%%%%%%%%%%%%%%%%%%%%%%%%%%%%%%%%%%%%%%%%%%%%%%%%%%%%%%%%%%%%%%%%

STS is a multidisciplinary construction on the intersection of dynamical systems and high-energy physics. It is interesting mainly because it unveils the physical essence of dynamical chaos. The central element in this theory is the topological supersymmetry breaking order parameter, which, as discussed here, is a field-theoretic embodiment of the butterfly effect. While this definitive feature of chaos is often viewed as a mere manifestation of the unpredictability of chaotic dynamics, STS highlights its other, more constructive side -- the dynamical memory -- and lays the foundation for formulating a consistent physical theory for it. Such theory could help address various important problems across diverse branches of modern science including neurodynamics, which is particularly intriguing considering the potentially significant role that neurodynamical BE may play in short-term memory and/or consciousness.

From a more general perspective, STS establishes a solid link between dynamical systems and high-energy physics theories. This link may help elevating fields such as hydrodynamics and neurodynamics to a higher level of mathematical precision, rigor, and predictive power. In return, high-energy physics can get access to a broad experimental testing ground for concepts that were previously confined solely to the realm of theoretical abstraction. 

\section*{Acknowledgements}
The author would like to acknowledge the initial support from DARPA BAA 'Physical Intelligence' and extend gratitude to Kang L. Wang, who played a pivotal role in enabling this work. Special thanks are also owed to Yurii E. Lozovik, Torsten A. Ensslin, Robert N. Schwartz, Massimiliano Di Ventra, Daniel Toker, Dmitri A. Riabtsev, Cheng-Zong Bai, and Eugene Ingerman, all of whom positively influenced the development of this work.

\section*{Declaration of competing interest}
The author declares no competing interests.

\section*{Data availability}
The code to reproduce data presented in Figs.(\ref{figure:2}), (\ref{figure:4}), and (\ref{figure:5}a) is available on request.

\bibliographystyle{unsrt}
\bibliography{main}

\begin{thebibliography}{10}

\bibitem{RevModPhys.57.617}
J.P. Eckmann and D.~Ruelle.
\newblock Ergodic theory of chaos and strange attractors.
\newblock {\em Rev. Mod. Phys.}, 57:617--656, 1985.

\bibitem{Chaos_book_1}
R.~Devaney.
\newblock {\em A First Course in Chaotic Dynamical Systems: Theory and
  Experiment}.
\newblock Addison-Wesley, 1992.
\newblock \url{https://doi.org/10.1201/9780429503481}.

\bibitem{Chaos_book_2017}
S.H. Strogatz.
\newblock {\em Nonlinear Dynamics and Chaos: With Applications to Physics,
  Biology, Chemistry, and Engineering (2nd ed.).}
\newblock CRC Press., 2015.

\bibitem{Gilmore}
R.~Gilmore.
\newblock Topological analysis of chaotic dynamical systems.
\newblock {\em Rev. Mod. Phys.}, 70:1455, 1998.

\bibitem{10.1063/5.0025924}
S.~Mangiarotti and C.~Letellier.
\newblock {Topological characterization of toroidal chaos: A branched manifold
  for the Deng toroidal attractor}.
\newblock {\em Chaos: An Interdisciplinary Journal of Nonlinear Science},
  31(1):013129, 01 2021.

\bibitem{Yorke_1975}
T.-Y. Li and J.~A. Yorke.
\newblock Period three implies chaos.
\newblock {\em The American Mathematical Monthly}, 82(10):985--992, 1975.

\bibitem{TransientChaos2015}
T.~Tél.
\newblock The joy of transient chaos.
\newblock {\em Chaos: An Interdisciplinary Journal of Nonlinear Science},
  25(9):097619, 07 2015.

\bibitem{Intermittency_Review}
S.~Elaskar and E.~del Río.
\newblock Review of chaotic intermittency.
\newblock {\em Symmetry}, 15(6), 2023.

\bibitem{Baxendale.10.1007/BFb0076851}
P.~H. Baxendale.
\newblock The lyapunov spectrum of a stochastic flow of diffeomorphisms.
\newblock In L.~Arnold and V.~Wihstutz, editors, {\em Lyapunov Exponents},
  pages 322--337, Berlin, Heidelberg, 1986. Springer Berlin Heidelberg.

\bibitem{Arnold.10.1007/BFb0076835}
L.~Arnold, W.~Kliemann, and E.~Oeljeklaus.
\newblock Lyapunov exponents of linear stochastic systems.
\newblock In Ludwig Arnold and Volker Wihstutz, editors, {\em Lyapunov
  Exponents}, pages 85--125, Berlin, Heidelberg, 1986. Springer Berlin
  Heidelberg.

\bibitem{Mot14}
A.E. Motter and D.K. Campbell.
\newblock Chaos at fifty.
\newblock {\em Phys. Today}, 66:27--33, 2013.

\bibitem{Rue14}
D.~Ruelle.
\newblock Early chaos theory.
\newblock {\em Phys. Today}, 67:9--10, 2014.

\bibitem{Poincare_celestial_dynamics}
H.~Poincar'e.
\newblock {\em New methods of celestial mechanics}, volume~13.
\newblock 1992.

\bibitem{ButterFly}
E.N. Lorenz.
\newblock Deterministic nonperiodic flow.
\newblock {\em J. Atmos. Sci.}, 20:130, 1963.

\bibitem{ARNOLD19831}
L.~Arnold and W.~Kliemann.
\newblock Qualitative theory of stochastic systems.
\newblock In A.T. Bharucha-Reid, editor, {\em Probabilistic Analysis and
  Related Topics}, pages 1--79. Academic Press, 1983.

\bibitem{Oks10}
B.~\"Oksendal.
\newblock {\em Stochastic Differential Equations: An Introduction with
  Applications}.
\newblock Springer, Berlin/Heidelberg, Germany, 2010.
\newblock \url{https://doi.org/10.1007/978-3-642-14394-6}.

\bibitem{LEJAN1984307}
Y.~Le~Jan and S.~Watanabe.
\newblock Stochastic flows of diffeomorphisms.
\newblock In K.~Itô, editor, {\em Stochastic Analysis}, volume~32 of {\em
  North-Holland Mathematical Library}, pages 307--332. Elsevier, 1984.

\bibitem{Kunita2019}
H.~Kunita.
\newblock {\em Stochastic Differential Equations and Stochastic Flows}, pages
  77--124.
\newblock Springer Singapore, Singapore, 2019.
\newblock \url{https://doi.org/10.1007/978-981-13-3801-4}.

\bibitem{Hairer_2001}
J.-P. Eckmann and M.~Hairer.
\newblock Invariant measures for stochastic partial differential equations in
  unbounded domains.
\newblock {\em Nonlinearity}, 14(1):133, 2001.

\bibitem{kupiainen2016renormalization}
A.~Kupiainen.
\newblock Renormalization group and stochastic pdes.
\newblock {\em Annales Henri Poincar\'e}, 17:497–535, 2016.
\newblock \url{https://doi.org/10.1007/s00023-015-0408-y}.

\bibitem{Heirer_2018}
M.~Hairer.
\newblock Renormalisation of parabolic stochastic pdes.
\newblock {\em Jpn. J. Math.}, 13:187–233, 2018.
\newblock \url{https://doi.org/10.1007/s11537-018-1742-x}.

\bibitem{Bedrossian_RecentReview}
J.~Bedrossian, A.~Blumenthal, and S.~Punshon-Smith.
\newblock A regularity method for lower bounds on the lyapunov exponent for
  stochastic differential equations., 2022.

\bibitem{Ovc11}
I.~V. Ovchinnikov.
\newblock Self-organized criticality as witten-type topological field theory
  with spontaneously broken becchi--rouet--stora--tyutin symmetry.
\newblock {\em Phys. Rev. E}, 83:051129, 2011.

\bibitem{OvcEntropy}
I.~V. Ovchinnikov.
\newblock Introduction to supersymmetric theory of stochastics.
\newblock {\em Entropy}, 18:108, 2016.

\bibitem{Torsten}
I.V. Ovchinnikov and T.A. En\ss{}lin.
\newblock Kinematic dynamo, supersymmetry breaking, and chaos.
\newblock {\em Phys. Rev. D}, 93:085023, 2016.

\bibitem{DMM_2}
M.~Di~Ventra and I.V. Ovchinnikov.
\newblock Digital memcomputing machines: From logic to dynmaics to topology.
\newblock {\em Annals of Physics}, 409:167935, 2019.

\bibitem{CG}
S.~Cecotti and L.~Girardello.
\newblock Stochastic and parastochastic aspects of supersymmetric functional
  measures: A~new non-perturbative approach to supersymmetry.
\newblock {\em Ann. Phys.}, 145:81--99, 1983.

\bibitem{DH}
I.T. Drummond and R.R. Horgan.
\newblock Stochastic processes, slaves and supersymmetry.
\newblock {\em J. Phys. A}, 45:095005, 2012.

\bibitem{Dijkgraaf}
R.~Dijkgraaf, D.~Orlando, and S.~Reffert.
\newblock Relating field theories via stochastic quantization.
\newblock {\em Nucl. Phys. B}, 824:365--386, 2010.

\bibitem{KS}
H.~Kleinert and S.V. Shabanov.
\newblock Supersymmetry in stochastic processes with higher-order time
  derivatives.
\newblock {\em Phys.~Lett.~A}, 235:105--112, 1997.

\bibitem{ZinnJustin}
J.~Zinn-Justin.
\newblock Renormalization and stochastic quantization.
\newblock {\em Nucl. Phys. B}, 275:135--159, 1986.

\bibitem{Lyapunov_SUSY}
R.~Graham.
\newblock Lyapunov exponents and supersymmetry of stochastic dynamical systems.
\newblock {\em EPL}, 5:101, 1988.

\bibitem{Witten_1982}
E.~Witten.
\newblock Supersymmetry and morse theory.
\newblock {\em J. Differ. Geom.}, 17:661--692, 1982.

\bibitem{Witten98}
E.~Witten.
\newblock Topological quantum field theory.
\newblock {\em Commun. Math. Phys.}, 117:353--386, 1988.

\bibitem{Witten981}
E.~Witten.
\newblock Topological sigma models.
\newblock {\em Commun. Math. Phys.}, 118:411--449, 1988.

\bibitem{Baulieu_1988}
L.~Baulieu and B.~Grossman.
\newblock A topological interpretation of stochastic quantization.
\newblock {\em Physics Letters B}, 212(3):351, 1988.

\bibitem{Baulieu_1989}
L.~Baulieu and M.~Singer.
\newblock The topological sigma model.
\newblock {\em Comms. in Math. Phys.}, 125(2):227, 1989.

\bibitem{labastida1989}
J.M.F. Labastida.
\newblock Morse theory interpretation of topological quantum field theories.
\newblock {\em Commun. Math. Phys.}, 123:641--658, 1989.

\bibitem{Blau}
M.~Blau.
\newblock The mathai-quillen formalism and topological field theory.
\newblock {\em Journal of Geometry and Physics}, 11(1):95 -- 127, 1993.

\bibitem{Brooks}
R.~Brooks, D.~Montano, and J.~Sonnenschein.
\newblock Gauge fixing and renormalization in topological quantum field theory.
\newblock {\em Phys. Lett. B}, 214:91, 1988.

\bibitem{TFT_BOOK}
D.~Birmingham, M.~Blau, M.~Rakowski, and G.~Thompson.
\newblock Topological field theory.
\newblock {\em Phys. Rep.}, 209:129, 1991.

\bibitem{Frenkel2007215}
E.~Frenkel, A.~Losev, and N.~Nekrasov.
\newblock Notes on instantons in topological field theory and beyond.
\newblock {\em Nucl.~Phys.~B}, 171:215--230, 2007.

\bibitem{Kible_10.1007/978-94-007-1029-0_1}
T.~W.~B. Kibble.
\newblock Symmetry breaking and defects.
\newblock In Henryk Arodz, Jacek Dziarmaga, and Wojciech~Hubert Zurek, editors,
  {\em Patterns of Symmetry Breaking}, pages 3--36, Dordrecht, 2003. Springer
  Netherlands.

\bibitem{ParSour}
G.~Parisi and N.~Sourlas.
\newblock Random magnetic fields, supersymmetry, and negative dimensions.
\newblock {\em Phys. Rev. Lett.}, 43:744--745, 1979.

\bibitem{ParSour1}
G.~Parisi and N.~Sourlas.
\newblock Supersymmetric field theories and stochastic differential equations.
\newblock {\em Nucl. Phys. B}, 206:321--332, 1982.

\bibitem{Gozzi2}
E.~Gozzi and M.~Reuter.
\newblock Algebraic characterization of ergodicity.
\newblock {\em Phys. Lett. B}, 233:383--392, 1989.

\bibitem{Deotto_1}
E.~Deotto, E.~Gozzi, and D.~Mauro.
\newblock Hilbert space structure in classical mechanics. i.
\newblock {\em J. Math. Phys.,}, 44:5902--5936, 2003.

\bibitem{Niemi1}
A.J. Niemi and P.~Pasanen.
\newblock Topological $\sigma$-model, hamiltonian dynamics and loop space
  lefschetz number.
\newblock {\em Phys. Letts. B}, 386:123--130, 1996.

\bibitem{Kurchan}
J.~Tailleur, S.~T\"anase-Nicola, and J.~Kurchan.
\newblock Kramers equation and supersymmetry.
\newblock {\em J. Stat. Phys.}, 122:557--595, 2006.

\bibitem{Nakahara}
M.~Nakahara.
\newblock {\em Geometry, Topology, and Physics}.
\newblock IOP Publishing, Bristol, UK, 1990.

\bibitem{chaos_2}
I.V. Ovchinnikov.
\newblock Topological field theory of dynamical systems. ii.
\newblock {\em Chaos}, 23(1):013108, 2013.

\bibitem{Rue02}
D.~Ruelle.
\newblock Dynamical zeta functions and transfer operators.
\newblock {\em Not. AMS}, 49:887--895, 2002.
\newblock \url{https://www.ams.org/journals/notices/200208/fea-ruelle.pdf}.

\bibitem{Chen_TFT_doi:10.1142/S0219887813500035}
W.~F. Chen.
\newblock Differential geometry from quantum field theory.
\newblock {\em International Journal of Geometric Methods in Modern Physics},
  10(04):1350003, 2013.

\bibitem{Slavik}
A.~Slavik.
\newblock Generalized differential equations: Differentiability of solutions
  with respect to initial conditions and parameters.
\newblock {\em Journal of Mathematical Analysis and Applications}, 402(1):261
  -- 274, 2013.

\bibitem{Mos02}
A.~Mostafazadeh.
\newblock Pseudo-supersymmetric quantum mechanics and isospectral
  pseudo-hermitian hamiltonians.
\newblock {\em Nucl. Phys. B}, 640:419--434, 2002.

\bibitem{Stochastic_PB_theorem}
X.~Zou, K.E. Wang, and D.~Fan.
\newblock Stochastic poincare-bendixson theorem and its application on
  stochasticc hoft bifurcation.
\newblock {\em International Journal of Bifurcation and Chaos}, 23(04):1350070,
  2013.

\bibitem{Ovchinnikov_2018}
I.~V. Ovchinnikov, Y.~Sun, T.~A. En{\ss}lin, and K.L. Wang.
\newblock Supersymmetric theory of stochastic {ABC} model.
\newblock {\em Journal of Physics Communications}, 2(6):065008, jun 2018.

\bibitem{Pesin}
Ya. Pesin.
\newblock Characteristic lyapunov exponents and smooth ergodic theory.
\newblock {\em Russ. Math. Surveys}, 32:55--114, 1977.

\bibitem{Review_Top_Entropy}
A.~Katok.
\newblock Fifty years of entropy in dynamics: 1958--2007.
\newblock {\em Journal of Modern Dynamics}, 1(4):545--596, 2007.

\bibitem{TopEntropy}
B.~Hasselblatt and A.~Katok.
\newblock {\em Handbook of Dynamical Systems}, volume~1A.
\newblock North–Holland, Amsterdam, 2002.

\bibitem{MirrorSymmetry}
K.~Hori, S.~Katz, A.~Klemm, R.~Pandharipande, R.~Thomas, C.~Vafa, R.~Vakil, and
  E.~Zaslow.
\newblock {\em Mirror Symmetry}.
\newblock American Mathematical Society and Clay Mathematics Institute,
  Cambridge, MA, USA, 2003.

\bibitem{DynSusyBrWitten}
E.~Witten.
\newblock Dynamical breaking of supersymmetry.
\newblock {\em Nucl. Phys. B}, 188:513--554, 1981.

\bibitem{Crutchfield}
J.P. Crutchfield.
\newblock Between order and chaos.
\newblock {\em Nature Physics}, 8:17–24, 2012.

\bibitem{CRUTCHFIELD200331}
J.~P. Crutchfield.
\newblock What lies between order and chaos?
\newblock In J.~Casti and A.~Karlqvist, editors, {\em Art and Complexity},
  pages 31--45. JAI, Amsterdam, 2003.

\bibitem{Bak87}
P.~Bak, C.~Tang, and K.~Wiesenfeld.
\newblock Self-organized criticality: An explanation of the $1/f$ noise.
\newblock {\em Phys. Rev. Lett.}, 59:381--384, 1987.

\bibitem{ChialvoLoh}
D.~R. Chialvo.
\newblock Emergent complex neural dynamics.
\newblock {\em Nat. Phys.}, 6(10):744--750, 2010.

\bibitem{SOC_controvercy}
N.W. Watkins, G.~Pruessner, S.C. Chapman, N.B. Crosby, and H.J. Jensen.
\newblock 25 years of self-organized criticality: Concepts and controversies.
\newblock {\em Space Science Reviews}, 198:3--44, 2016.

\bibitem{beggs2004}
J.M. Beggs and D.~Plenz.
\newblock Neuronal avalanches are diverse and precise activity patterns that
  are stable for many hours in cortical slice cultures.
\newblock {\em J. Neurosci.}, 24(22):5216--5229, 2004.

\bibitem{Levina1}
A.~Levina, J.~M. Herrmann, and T.~Geisel.
\newblock Dynamical synapses causing self-organized criticality in neural
  networks.
\newblock {\em Nat Phys}, 3(12):857--860, 2007.

\bibitem{sine_Gordon_Kink_Bath}
N.~R. Quintero, A.~S\'anchez, and F.~G. Mertens.
\newblock Overdamped sine-gordon kink in a thermal bath.
\newblock {\em Phys. Rev. E}, 60:222--230, Jul 1999.

\bibitem{Classics}
G.~Eilenberger.
\newblock Bremsstrahlung from solitons.
\newblock {\em Zeitschrift fur Physik B - Condensed Matter}, 27(2):199--203,
  1977.

\bibitem{Josephson_general}
P.~S. Lomdahl, O.~H. Soerensen, and P.~L. Christiansen.
\newblock Soliton excitations in josephson tunnel junctions.
\newblock {\em Phys. Rev. B}, 25:5737--5748, May 1982.

\bibitem{PhysRevB.54.1234}
Z.~Hermon, E.~Ben-Jacob, and G.~Sch\"on.
\newblock Charge solitons in one-dimensional arrays of serially coupled
  josephson junctions.
\newblock {\em Phys. Rev. B}, 54:1234--1245, Jul 1996.

\bibitem{Review_Frenkel_Kontorova}
O.M. Braun and Y.S. Kivshar.
\newblock Nonlinear dynamics of the frenkel–kontorova model.
\newblock {\em Physics Reports}, 306(1–2):1 -- 108, 1998.

\bibitem{ovchinnikov2020}
I.V. Ovchinnikov, W.~Li, Y.~Sun, A.~E. Hudson, K.~Meier, R.~N. Schwartz, and
  K.~L. Wang.
\newblock Criticality or supersymmetry breaking?
\newblock {\em Symmetry}, 12(5):805, 2020.

\bibitem{NeuronalDynamicsBook}
W.~Gerstner, W.~M. Kistler, R.~Naud, and L.~Paninski.
\newblock {\em Neuronal Dynamics: From Single Neurons to Networks and Models of
  Cognition}.
\newblock Cambridge University Press, Cambridge, 2014.
\newblock \url{https://neuronaldynamics.epfl.ch/online/index.html}.

\bibitem{Izhikevich}
E.M. Izhikevich.
\newblock {\em Dynamical Systems in Neuroscience}.
\newblock MIT Press, Cambridge, 2007.
\newblock \url{https://doi.org/10.7551/mitpress/2526.001.0001}.

\bibitem{li2018}
W.~Li, I.V. Ovchinnikov, H.~Chen, Z.~Wang, A.~Lee, H.~Lee, C.~Cepeda, R.N.
  Schwartz, Karlheinz M., and K.L. Wang.
\newblock A basic phase diagram of neuronal dynamics.
\newblock {\em Neural Comput.}, 30(9):2418--2438, 2018.

\bibitem{Kang}
I.~V. Ovchinnikov, R.N. Schwartz, and K.L. Wang.
\newblock Topological supersymmetry breaking: The definition and stochastic
  generalization of chaos and the limit of applicability of statistics.
\newblock {\em Modern Physics Letters B}, 30(08):1650086, 2016.

\bibitem{toker2020}
D.~Toker, F.T. Sommer, and M.~D’Esposito.
\newblock A simple method for detecting chaos in nature.
\newblock {\em Comm. Biol.}, 3:11, 2020.

\bibitem{Kopietz2010}
P.~Kopietz, L.~Bartosch, and F.~Sch{\"u}tz.
\newblock {\em Wilsonian Renormalization Group}, pages 53--89.
\newblock Springer Berlin Heidelberg, Berlin, Heidelberg, 2010.

\bibitem{SusyReview_2014}
Y.~Shadmi.
\newblock Introduction to supersymmetry.
\newblock In M.~Mulders and G.~Zanderighi, editors, {\em Proceedings of the
  2014 European School of High-Energy Physics}, volume~3 of {\em CERN Yellow
  Reports}, pages 95--123. Garderen, the Netherlands, 2016.
\newblock \url{https://doi.org/10.5170/CERN-2016-003.95}.

\bibitem{topological_order_review}
K.~J. Satzinger, Y.-J Liu, A.~Smith, and et.al.
\newblock Realizing topologically ordered states on a quantum processor.
\newblock {\em Science}, 374(6572):1237--1241, 2021.

\bibitem{Yamada_Ohkitani_doi:10.1143/JPSJ.56.4210}
M.~Yamada and K.~Ohkitani.
\newblock Lyapunov spectrum of a chaotic model of three-dimensional turbulence.
\newblock {\em Journal of the Physical Society of Japan}, 56(12):4210--4213,
  1987.

\bibitem{CFT_SOC}
M.~N. Najafi, J.~Cheraghalizadeh, and H.~J. Herrmann.
\newblock Self-organized criticality in cumulus clouds.
\newblock {\em Phys. Rev. E}, 103:052106, May 2021.

\bibitem{AdSCFT_book}
M.~Natsuume.
\newblock Ads/cft duality user guide.
\newblock {\em Lecture Notes in Physics}, 903, 2015.

\bibitem{holography}
H.C. Longuet-Higgins.
\newblock Holographic model of temporal recall.
\newblock {\em Nature}, 217, 1968.

\bibitem{Kohonen1989}
T.~Kohonen.
\newblock {\em Optical Associative Memories}, pages 269--284.
\newblock Springer Berlin Heidelberg, Berlin, Heidelberg, 1988.

\bibitem{Sommer2012}
F.T. Sommer.
\newblock {\em Associative Memory and Learning}, pages 340--342.
\newblock Springer US, Boston, MA, 2012.

\end{thebibliography}

\end{document}